\documentclass[conference]{IEEEtran}
\IEEEoverridecommandlockouts

\newcommand{\conftitle}{Submitted to 12th International Workshop on Science Gateways (IWSG 2020), 10-12 June 2020}

\usepackage{fancyhdr}
\fancypagestyle{pageStyle}{%
    \fancyhf{}
    \fancyhead[C]{\conftitle}

}


\usepackage{fancyhdr}
\usepackage{footnote}
\usepackage[hidelinks]{hyperref}
\usepackage{listings}
\usepackage{xcolor}

\usepackage{graphicx}
\usepackage{float}

\colorlet{punct}{red!60!black}
\definecolor{background}{HTML}{EEEEEE}
\definecolor{delim}{RGB}{20,105,176}
\colorlet{numb}{magenta!60!black}

\definecolor{eclipseStrings}{RGB}{42,0.0,255}
\definecolor{eclipseKeywords}{RGB}{127,0,85}
\colorlet{numb}{magenta!60!black}

\lstdefinelanguage{json}{
    basicstyle=\normalfont\ttfamily,
    commentstyle=\color{eclipseStrings}, 
    stringstyle=\color{eclipseKeywords}, 
    showstringspaces=false,
    breaklines=true,
    string=[s]{"}{"},
    comment=[l]{:\ "},
    morecomment=[l]{:"},
    literate=
        *{0}{{{\color{numb}0}}}{1}
         {1}{{{\color{numb}1}}}{1}
         {2}{{{\color{numb}2}}}{1}
         {3}{{{\color{numb}3}}}{1}
         {4}{{{\color{numb}4}}}{1}
         {5}{{{\color{numb}5}}}{1}
         {6}{{{\color{numb}6}}}{1}
         {7}{{{\color{numb}7}}}{1}
         {8}{{{\color{numb}8}}}{1}
         {9}{{{\color{numb}9}}}{1}
}

\lstdefinelanguage{jsonX}{
    basicstyle=\normalfont\ttfamily,
    showstringspaces=false,
    breaklines=true,
    frame=lines,
    backgroundcolor=\color{background},
    literate=
     *{0}{{{\color{numb}0}}}{1}
      {1}{{{\color{numb}1}}}{1}
      {2}{{{\color{numb}2}}}{1}
      {3}{{{\color{numb}3}}}{1}
      {4}{{{\color{numb}4}}}{1}
      {5}{{{\color{numb}5}}}{1}
      {6}{{{\color{numb}6}}}{1}
      {7}{{{\color{numb}7}}}{1}
      {8}{{{\color{numb}8}}}{1}
      {9}{{{\color{numb}9}}}{1}
      {\{}{{{\color{delim}{\{}}}}{1}
      {\}}{{{\color{delim}{\}}}}}{1}
      {[}{{{\color{delim}{[}}}}{1}
      {]}{{{\color{delim}{]}}}}{1},
}

\fancypagestyle{pageStyle}{%
    \fancyhf{}
    \fancyhead[C]{}

}


\ifCLASSINFOpdf
\usepackage{graphicx}
\else
\fi
\usepackage{url}

\begin{document}

%
\title{Certifying Provenance of Scientific Datasets with Self-sovereign Identity and Verifiable Credentials}
\renewcommand\IEEEkeywordsname{Keywords}

\author{\IEEEauthorblockN{Iain Barclay, Alun Preece, Ian Taylor}
\IEEEauthorblockA{Crime and Security Research Institute,\\Cardiff University,\\Cardiff, UK\\
Email: BarclayIS@cardiff.ac.uk}

\and
\IEEEauthorblockN{Swapna Radha, Jarek Nabrzyski}
\IEEEauthorblockA{Center for Research Computing,\\University of Notre Dame,\\Notre Dame, IN, USA\\Email: sradha@nd.edu}
}


%



\maketitle
\thispagestyle{pageStyle}
\pagestyle{fancy}
\renewcommand{\headrulewidth}{0pt} 

\begin{abstract}
In order to increase the value of scientific datasets and improve research outcomes, it is important that only trustworthy data is used. This paper presents mechanisms by which scientists and the organisations they represent can certify the authenticity of characteristics and provenance of any datasets they publish so that secondary users can inspect and gain confidence in the qualities of data they source. By drawing on data models and protocols used to provide self-sovereign ownership of identity and personal data to individuals, we conclude that providing self-sovereignty to digital assets offers a viable approach for institutions to certify qualities of their datasets in a cryptographically secure manner, and enables secondary data users to efficiently perform verification of the authenticity of such certifications. By building upon emerging standards for decentralized identification and cryptographically verifiable credentials, we envisage an infrastructure of tools being developed to foster adoption of metadata certification schemes, and improving the quality of information provided in support of shared data assets.
\end{abstract}




%
\IEEEpeerreviewmaketitle

\section{Introduction}

Owners and publishers of scientific datasets and digital assets have an opportunity to increase the scientific value~\cite{gil2016toward} of their dataset or further the commercial opportunities~\cite{miller2013data} for other types of digital asset by demonstrating the faithfulness of claims about their data, or the authenticity of their assets, through mechanisms which can act as digital watermarks~\cite{acken1998watermarking}. In order to have confidence in the quality and suitability of a dataset for their needs, potential users and other stakeholders need to have trust in claims made by the originators of the dataset. In practice, even in environments where funding organisations insist on researchers sharing data, there can be a resistance to reuse as Pisani et al.~\cite{pisani2018} found ``lower-than-expected reuse of shared data may be because potential secondary users have few ways of checking the quality of those data''. As such, a scheme which allows potential users to access dataset properties and attributes in the form of signed credentials, with an assurance that the credentials related to the data that they were inspecting, were issued by an authorised party and had not been revoked, or tampered with since issuance has the potential to increase the use of shared datasets.
 
This paper discusses methods for providing signed certifications relating to properties or attributes of digital assets. The discussion considers a scientific dataset, and reviews mechanisms by which parties with ownership or authority over the dataset can attach digital credentials to the dataset so that prospective users and other interested parties can access and verify the credentials to gain confidence in the dataset's qualities. The credential used to illustrate the discussion is an assertion from the originators of a hypothetical dataset that the data it contains has been ethically sourced and is cleared for further use --- a need identified by Scott et al.~\cite{scott2019owns} when considering the implications of recorded voices in a published dataset being used in different scenarios. 

The paper continues as follows: Section~\ref{sec:current} introduces current mechanisms for describing properties and metadata about shared datasets, and provisions for evidencing attributes of datasets to provide provenance. Section~\ref{sec:digital} discusses established methods for secure information publishing using public and private key signatures, and identifies shortcomings in these methods when applied to sharing assertions about shared digital assets and datasets. The proposed approach uses Self-sovereign Identity (SSI)~\cite{muhle2018survey}, and the enabling concepts of decentralised identifiers (DIDs)~\cite{hughes2019} and verifiable credentials (VCs)~\cite{sporny2018} which are introduced in Section~\ref{ssiconcepts}. Section \ref{usingvcs} explains how these SSI data models and protocols can be adapted to provide a mechanism that allows the publisher of a dataset to issue credentials attesting that their dataset has certain properties. Section~\ref{usingssi} extends the discussion to consider automation of credential exchange by treating data assets as self-sovereign entities, with software agents mediating the presentation and verification of credentials between publishers and potential dataset users. The paper concludes with a discussion of our ongoing work in this area as we seek to further integrate existing data sharing schemes with verifiable credentials so that they can be used together to provide improved confidence and trust in shared scientific datasets.

\section{Sharing Datasets, Attributes and Provenance}
\label{sec:current}
Petabytes of data are generated every day by various scientific facilities ranging from telescopes to the world’s largest machine viz. Hadron Collider~\cite{Simons2012Implementing}. However, due to a lack of proven methods to assess the quality of data and inadequate archival processes, secondary users can effectively utilize only a small subset of the collected data. For the data to be useful to its full extent it should be verifiable for its origin and trustworthiness~\cite{hills2015}, especially when it is applied in the field of scientific research.

Methods are available to add unique attributes to shared datasets, which add a certain level of confidence to the quality of data. Such methods are briefly explained in the following section. The Digital Object Identifier (DOI) System is widely used for identification and management of intellectual content and metadata, and to connect end-users with the requested content. It provides a technical and social framework to identify data by using persistent, interoperable identifiers that can be used across digital networks~\cite{Simons2012Implementing}. Within dataset records located by DOIs, different domains have their own specifications and requirements for metadata. Earth Science Information Partners, founded by NASA, recommends the use of the Provenance and Context Content Standard (PCCS) matrix to perform identification, capturing and tracking of all metadata that can be used to validate the data and to facilitate efficient scientific reproducibility~\cite{hills2015}. NASA has use of the PCCS matrix as a requirement for all new Earth science missions, using it to capture and record metadata such as dataset product documentation, dataset product validation which includes the validation record and datasets, dataset calibration information such as calibration method, data, and software used. Additionally, the OpenGIS Sensor Model Language (SensorML) Encoding standard is used to demonstrate various important attributes of sensor and sensory systems such as geometric, dynamic and observational characteristics~\cite{ramapriyan2011nasa}.

In the Biometrics field, Czajka et al.~\cite{czajka2016verification} introduce a digital watermarking method that can prove the trustworthiness of iris images without requiring to add supporting data to the iris image. This watermarking based method was designed to certify that biometric data had originated from a genuine sensor. Zhuang et al.~\cite{zhuang2020facial} discuss the experimentation and evaluation of some of the existing feature extraction methods that are used to measure facial weaknesses. Since an open source annotated facial weakness images dataset was unavailable,  experimentation involved first creating a facial weakness dataset using images and videos from public repositories like Google Images and YouTube. The generated ``neurologist-certified dataset'' was subjected to multiple reviews by experienced neurology trainees and also experts in the field of neurology. If the dataset created during this study is to be effectively shared and applied in future experiments, there would be value in providing demonstrable proof that it was reviewed and certified by neurology experts.

Existing methods used to provide provenance and demonstrate the trustworthiness of datasets are designed to cater to the need of specific use cases or domains. Although these methods are being successfully implemented in their respective fields, there is a lack of a common framework or verification mechanism that can be seamlessly applied across all areas of research and other fields. In order to foster the development of such a framework, and to facilitate the provision of high-quality toolsets, it is proposed to provide a mechanism by which datasets and core attributes of metadata about those datasets can be certified and verified with cryptographic assurance using open standards-based models and protocols.

\section{Certification of Digital Assets}
\label{sec:digital}
\begin{figure}[ht]
\centering
\includegraphics[width=0.45\textwidth]{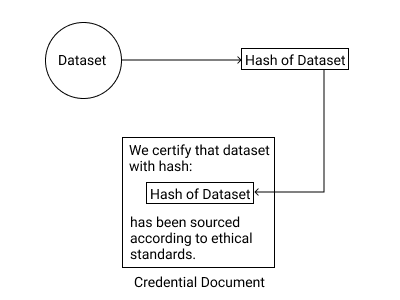} 
\caption{A Credential Document containing a cryptographic hash of a dataset.}
\label{fig:hash}
\end{figure}

When evaluating a certification or attestation about an entity, it is critical that the certificate can be inextricably linked to the entity which it describes, otherwise the certificate is of limited value --- a photographic identity card, for example, is only of use in attesting the identity of its holder if the inspector of the identity card believes that the photograph on the card provides a visual match to the holder. For digital assets, it is common to use a cryptographic hash value~\cite{preneel1994cryptographic} derived from the asset itself to provide a unique digital fingerprint of the asset, and it is proposed to apply this technique to the identification of the datasets discussed in this study. A cryptographic hash is a unique fixed length value that can be generated for a dataset and which will change if the data in the dataset changes in any way, even by a single bit. If the hash value is recorded inside a certificate or credential document, then it can be assured that if the cryptographic hash of the dataset under inspection matches the stored cryptographic hash, then the credential refers to the same dataset. For example, if the certificate states \emph{The dataset with the cryptographic hash 0x1122EEFF has been sourced ethically} then it will uniquely identify the dataset to which it refers. Note that cryptographic hash values are typically 256 or 512 bits in length~\cite{gueron2011sha}, so the structure of the credential would not likely be readable in sentence form, but the cryptographic hash would be embedded in the credential in some way, as Figure~\ref{fig:hash} illustrates.

\begin{figure} [ht]
\centering
\includegraphics[width=0.45\textwidth]{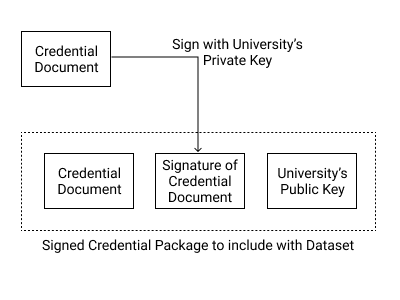} 
\caption{Signed Credential Package for a dataset instance.}
\label{fig:signing}
\end{figure}

Consider the case in which the information in the dataset's credential is public information, and can be accessed freely by anyone. In this case, an authorised representative of the publisher of the dataset might create a credential document asserting that the dataset represented by the supplied cryptographic hash has been ethically sourced and make a digital signature of the document before finally storing the credential document, the signature and the corresponding public key~\cite{rivest1978method} in a place where it can be inspected (for example, as part of the supporting material for the shared dataset~\cite{pepe2014astronomers}), as illustrated in Figure~\ref{fig:signing}. Anyone subsequently accessing the dataset's distribution archive can use the public key to verify that the credential document has been signed by the publisher. As discussed above, the credential document would need to include a statement along the lines of \emph{The dataset with the cryptographic hash 0x1122EEFF has been sourced ethically} in order to provide an intrinsic link between the credential document and the dataset it ratifies.

The primary advantage of such a system is that it provides assurance that the certification document is genuine and has not been tampered with, and the system relatively simple to implement as it makes use of well established cryptographic techniques. The publisher will have to establish a process for issuing and managing keys, and in particular for safeguarding access to the private key to ensure that only authorised representatives of the publisher can access and use it.

There are, however, shortcomings in this scheme: most notably, a signed credential issued alongside the dataset cannot be withdrawn or modified if a released dataset is subsequently found to have issues with its ethical status --- previously-downloaded copies of the dataset will continue to be accompanied by the credential asserting that the data is ethically sourced, even if evidence subsequently disproves this. Furthermore, including the public key inside the distribution package doesn't guarantee that the public key belongs to the claimed signing party --- a malicious party could simply build a new distribution, including a fake certificate signed with their own key, and include that key in the distribution package, giving the appearance that everything was in order whilst propagating false claims about the dataset. This could be mitigated to some degree by hosting the public key on a website or other location that was known to be under the control of the publishing party and then sharing the public key's location over a trusted channel. As such, the informal digital signature-based scheme presented here for verifying a certification document would benefit from having well-known mechanisms and structure for sharing public key locations, and for expressing semantic information about the dataset, such that processes for signing and metadata verification could be automated and be more readily interoperable with other systems and processes.

\section{Introducing Self-Sovereign Identity }
\label{ssiconcepts}
The term Self-sovereign Identity (SSI)~\cite{allen2016path} is used to describe the ability of an individual to take ownership of their personal data and to control who has access to that data, without the need for centralized infrastructure. For instance, in a traditional siloed model identity system, a single or group of identity providers administer the entire system and usually maintains a centralized repository of user identities. This makes it challenging for users to have control over their own data and they depend on administrators for distribution and verification of data. Another disadvantage of a siloed model identity system is that it is not portable. The portability issue of the siloed model is solved by the federated model, which allows some degree of portability by allowing different services to share details about a single user; but the user still does not have complete ownership of their identity. Self-sovereign identity solves the issue of ownership by allowing users to maintain control over their identity. With the use of self-sovereign identity, it is possible to secure users’ identity information from unauthorized disclosure by allowing the identity owner to selectively disclose data based on the requirement from the verifier. The data owner thus has the ability to decide how their identity and their personal data should be used and who has access to it.

The SSI community have developed data models and protocols~\cite{sporny2018} that provide cryptographically verifiable mechanisms for validating identities and issuing and presenting proofs of credentials. These mechanisms build on distributed ledger technologies and linked JSON-LD~\cite{lanthaler2012using} data documents to provide immutable proof of control of an identifier, and to facilitate the secure exchange of credentials based upon these identifiers between consenting parties.

Whilst a significant focus of effort in the SSI community has been on personal identity and data privacy for individuals, the underlying computer science techniques can be applied to any type of entity, including digital assets and scientific datasets, where the attention of this paper focuses on demonstrating that SSI techniques can be used to provide assertions about the qualities and provenance of a dataset.

\section{Representing Attributes as Verifiable Credentials}
\label{usingvcs}
A core tenet of the decentralised identifier model is that parties claiming to be the controller of a DID can provide cryptographic proof that this is the case, facilitated by a protocol that the DID provides a route to a verification mechanism. This route is typically provided in the form of a JSON-LD document containing the public key of the DID, along with the methods by which a party can verify. By using the published verification mechanisms the holder of a document allegedly signed by the DID’s controller can obtain cryptographic proof that it was indeed signed by the DID controller, and furthermore can verify that the document has not been tampered with since it was signed. 

This prescriptive mechanism for a party to prove that they have access to the private keys relating to a DID is utilised when issuing Verifiable Credentials (VC), which can be as simple as a document claiming a DID. If the claim document is signed by a reputable and trusted party, and the DID of that party is known, then the claims in the document can be taken to have been issued by the trusted party. In other words, if a university signs a document using the private key of a DID they control, then the claims in the document can be taken to be claims that the university is willing to endorse. The credential issuer will be the trust anchor in the system, such that anyone relying on credentials provided by the issuer will need to have trust in the issuer themselves to place value on the credentials~\cite{linn2000trust}. Where the issuer is a university, NGO or other well-regarded organisation this trust may be inherent, in other cases the issuer may need to source credentials from bodies with a better established reputation in order to assert their own qualities as a trustworthy issuer of credentials.

Providing a mechanism to assure verifying parties that a DID belongs to a known and trusted authority is a governance challenge, requiring both policies and infrastructure to provide a white list or other trustworthy records that can be checked. In the short term, a DID scheme has been proposed which makes use of the fact that most organisations run web sites, and typically have certificates proving the legitimacy of the identity of the web site. The \textit{did:web}~\cite{Terbu2020} scheme takes advantage of this by utilising the web site of an organisation to host the DID document, resolving \textit{did:web} to a JSON-LD file located on the web site with a well known path~\cite{nottingham2010} and relying on only authorised users being able to upload files to an organization's official web site.

\noindent
\begin{minipage}{\linewidth}
\begin{lstlisting}[language=json, label={lst:uniofsci}, caption={A fragment of the UniOfScience DID Document},captionpos=b]
{
    "@context": "https://w3id.org/did/v1",
    "id": "did:web:uniofscience.com",
    "authentication": [{
        "id": "did:web:uniofscience.com",
        "type": "Ed25519VerificationKey2018",
        "controller": "did:web:uniofscience.com",
        "publicKeyBase58": "71ANMccQC..."
      }]
      ...
}
\end{lstlisting}
\end{minipage}

An open source software package \textit{vc-js}~\cite{2020vcjs} can be used to generate VC documents based on DIDs using many schemes, including \textit{did:web}. In order to produce a VC document for an illustrative dataset, domain names were registered for \textit{UniOfScience}, a fictitious university, and to represent a web site for the dataset, at \textit{DIDdoi.com}, and DID Documents were crafted for each, such that resolving the did:web address through the published route would reach the appropriate DID Document. Listing~\ref{lst:uniofsci} shows part of the DID Document for \textit{UniOfScience}. 

The second required component is a Credential Schema~\cite{sporny2019}, which defines the semantic vocabulary to be used to describe the attributes of the dataset and provides the format in which the claims about a particular subject will be made. To produce a VC document the \textit{vc-js} library was integrated with the Node.js Express~\cite{hahn2016express} framework to enable a simple web form to be served to allow a user --- perhaps a scientist preparing to publish a dataset --- to enter information about the dataset which was subsequently used to populate data fields in the credential schema, and \textit{vc-js} invoked to encapsulate these values in a Verifiable Credential JSON-LD document containing a proof issued by the DID belonging to \textit{UniOfScience}.

The inclusion of the DID of the issuer (and signatory) of the VC document enables other parties to verify its state, which is achieved by resolving the DID to locate the DID Document holding descriptors of the mechanisms for checking signatures, usually by provision of the public key. Verifiers can use methods in \textit{vc-js} to receive cryptographic proof of the authenticity of the VC document, assuring them that it hasn’t been tampered with since it was issued. As the payload of the VC document contains the DID of the dataset that it refers to, verifiers have cryptographic proof that the issuer has signed a document attesting to the properties of the DID of the subject. If the DID relates to a dataset, then the verifier can be assured that the VC document carries signed assurances about the properties of the dataset.

Presentation of a VC document provides systematic improvements over an ad hoc digitally signed document, through the publication of the location of the public key of the subjects and the use of a semantic vocabulary for expressing metadata claims, but still exhibits shortcomings. Among these is the possibility for anyone who holds a copy of the VC document to present it, even though it may no longer be relevant or valid. If this is done with ill-intent it could be considered a reply attack~\cite{nist2017}. A mechanism to prevent such replay attacks is for the verifier of a credential claim to ask the holder to present a Verifiable Proof, which takes the form of a JSON-LD document signed by the VC holder containing a challenge, typically a nonce, issued by the verifier along with the credential and its proof. By inspection of the Verifiable Proof document through resolution of the DID of its issuer, and comparing this DID with the DID of the VC's subject, the verifier can determine that the challenge response is acceptable, and that the holder of the VC is still content to present it. Further verification of the credential contained in the Verifiable Proof can demonstrate that the credential has not been tampered with, and thereby provide assurance that a valid credential about the dataset has been issued by an authorised party, to a holder who still considers the credential appropriate to share.

\section{Self-sovereignty of Digital Assets}
\label{usingssi}
The process described thus far has implied that parties wishing to verify credentials make requests and human operators are on hand to manage private keys and to create and sign Verifiable Proof documents, which would quickly become impractical where there was high demand or a need for a timely response. By considering the scientific dataset as a self-sovereign entity in its own right, with control over its own credentials, we can begin to automate these processes using a construct, exemplified by the open source Hyperledger Aries project~\cite{Aries} and Evernym's commercial application libvcx\footnote{\url{https://www.evernym.com}}, whereby the entity is represented by a software application (termed an agent) operating on behalf of the entity and mediating access to the entity's credentials. In the scenario presented, a cloud-based software agent (DSA) will represent the published dataset (DS) and will provide mechanisms to generate and store the private keys and issued VC documents securely in a digital wallet. Parties with interests in DS will interact with the representing agent, DSA, by addressing it through its publicly shared decentralised identifier (DID).

The dataset publisher will also have a software agent representing their role in the transactions, with the capability to issue credentials. The publisher's agent will be the trust anchor in the system, such that anyone relying on data credentials provided by the publisher's agent will need to have trust in the publisher themselves in order to place value on the credentials, as in the \textit{did:web} scheme described above. For the discussion, the publisher is a university that employs the scientist who seeks to share a dataset with the research community.

\begin{figure} [th]
\centering
\includegraphics[width=0.45\textwidth]{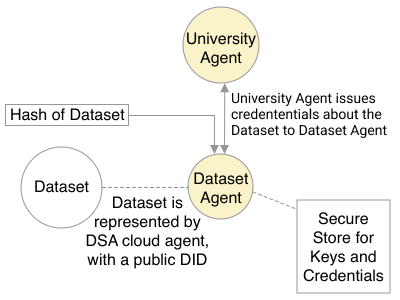} 
\caption{A publisher agent (University Agent) interacts and issues credentials to a dataset agent.}
\label{fig:did}
\end{figure}

In the first instance the university will configure and launch a new software process to act as the agent DSA and represent DS in future transactions. This software process will be part of the infrastructure provided by an implementation of the DID protocols, such as the ACA-Py~\cite{AcaPy} cloud agent component from the Hyperledger Aries project. As part of the commissioning procedure for the agent, a configuration script will be run to generate a digital wallet for the agent, which will hold its private keys and credentials, and generate a decentralised identifier (DID), by which DSA will be addressed. 

The university will then establish a secure connection between their publisher agent and DSA, which will be manifested using DID protocols, and the resultant connection will be used to issue the appropriate credentials to DSA, as shown in Figure~\ref{fig:did}. Credentials are structured according to published JSON-formatted Credential Definition schema~\cite{lux2019full} and contain a set of key-value pairs which the issuing party asserts are true about the holding entity. For example, in Listing~\ref{lst:hash}, \textit{Hash of Data} is a credential holding the cryptographic hash of the dataset, which inextricably links the credential to the dataset it represents, and \textit{Data Ethically Sourced} represents the ethical status of the dataset, as stated by the publisher. A practical scheme will hold other credentials, and could include a credential expiry date or other conditions for credential usage. 

\noindent
\begin{minipage}{\linewidth}
\begin{lstlisting}[language=json, label={lst:hash}, caption={A sample credential set},captionpos=b]
{
    "Hash of Data": "0xFFEE...AA1122",
    "Data Ethically Sourced": "YES"
}
\end{lstlisting}
\end{minipage}

\begin{figure} [th]
\centering
\includegraphics[width=0.45\textwidth]{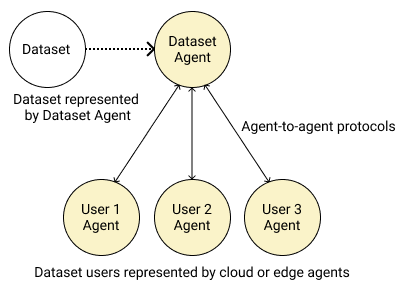} 
\caption{Agents represent parties in the relationship.}
\label{fig:agents}
\end{figure}

The DID for the dataset agent, DSA, is a public address, analogous to a website address, which should be shared in the downloadable package for the Dataset DS that it represents, such that users can use this address to request proof of the credentials of DS. Users of DS will themselves use software agents to communicate with its agent, DSA, these may be edge agents stored on a mobile device or cloud agents hosted by a web service. Any user with knowledge of the DID for DSA can seek to establish a connection with DSA and to request proof of the dataset's credentials via their own agent, and if the system policies permit, DSA will respond, presenting proof of the credentials it holds. These credentials will include the cryptographic hash of the dataset, to provide an inextricable link to the underlying dataset it represents, along with its ethical sourcing status, as written in the credential by the publisher. The public DID of the university will be included in the returned credential proof, and can be used as a trust anchor to assure that the credentials originated from the university. DID protocols ensure that the returned proof is cryptographically provable to have originated from the issuing university and to have not expired, been revoked or tampered with in any way. The architecture for such a scheme is shown in Figure~\ref{fig:agents}. An implementation of this architecture has been prototyped through access to a commercial software platform provided for evaluation by Evernym, and  demonstrated to work as designed.

\section{Conclusions and Future Work}
\label{sec:conclusion}
This paper has explored the use of emerging protocols and data models for decentralised identifiers and verifiable credentials, as they can be applied to providing certification of qualities and attributes of digital assets. The discussion has been motivated through the use of a hypothetical ethical certification applied to a scientific dataset, and has described how decentralised identifiers and verifiable credentials can be used to provide users of a dataset with access to a well-structured and semantically-rich digital attestation of qualities of the dataset.

By building upon concepts and data models from Self-sovereign Identity research, it is hoped that a standard's based approach to credential sharing and verification of properties of digital assets will emerge, such that users will be able to leverage a range of interoperable tools and platforms both to certify and to verify digital assets, whilst taking advantage of research into the application of advanced techniques including selective disclosure and zero knowledge proofs to further the security and privacy preserving capabilities of digital credentials. 

Continuation of the work presented here will look towards further implementation of the architectures discussed, and in particular how the proposed approach can be combined with the widely adopted DOI scheme in order to provide definitive assurance of attributes and qualities to users of datasets across scientific fields.

\bibliographystyle{IEEEtran}
\bibliography{IEEEabrv,datasharing}

\end{document}